# Observation of Quantized Conductance in the Surface Bands of Bismuth Nanowires

T. E. Huber,[1*] S. Johnson,[1] L. Konopko,[2,3] A. Nikolaeva,[2,3] A. Tsurkan[2] and M. J. Graf [4]

[1] Howard University, Washington, DC 20059, USA

[2] Academy of Sciences, Chisinau, Moldova, MD-2028

[3] International Laboratory of High Magnetic Fields and Low Temperatures, 53-421 Wroclaw, Poland

[4] Department of Physics, Boston College, Chestnut Hill, MA 02467, USA

We report the experimental observation of quantized conductance by Rashba bands at the surface of 50-nm Bi nanowires. With increasing magnetic fields along the wire axis, the wires exhibit a stepwise increase of conductance with as many as four distinct plateaus together with oscillatory thermopower. The observations can be accounted for by the increase of the number of propagating surface modes. The modes have very high mobility and are associated with Aharonov-Bohm interference around the perimeter consistently with theory of quasiballistic one-dimensional nanowires considering that Lorentz forces decouple the modes from scattering at the nanowire surface.

Surface bands, which appear as a result of surface spin-orbit coupling (SOC) [1-3], represent a new direction in the field of two-dimensional (2D) electron gases. SOC surface bands have been observed in many materials, including Bi and topological insulators (TIs) like $Bi_2Se_3$ and $Bi_2Te_3$, via angle resolved photoemission spectroscopy (ARPES). The behavior of these bands varies from the well-studied cases of semiconductor interfaces and graphene [4] because the electron motion is uniquely spin polarized. However, a very important feature of 2D electron gases is very high electronic mobility and quantum-mechanically enhanced electronic transport, and these properties have not been demonstrated with SOC bands. Here we show that the SOC bands in 50-nm Bi nanowires exhibit stepwise conductance and achieve very high mobilities.

Magnetoresistance (*MR*) studies of small diameter Bi nanowires [5,6], nanoribbons [7] and TI nanoribbons [8-10] clearly demonstrate surface conduction. With an applied magnetic field *B* parallel to the wire axis, the surface electronic wave function picks up the Berry and Aharonov-Bohm (AB) phase of $2\pi\phi/\phi_o$, with $\phi$ equaling the total magnetic flux through the cross sectional area and $\phi_0 = h/e$, the flux quantum, causing the *MR* to exhibit oscillations with *h/e* and *h/2e* flux periods due to quantum mechanical interference [11-13]. Although the presence of *h/2e* periods is suggestive of strong disorder where the mean free path (*mfp*) is less than the diameter, the flux dependent subband structure giving rise to *h/e* and *h/2e* periods is also a characteristic of spin-split high mobility, quasiballistic, surface bands in semiconductor nanowires [14-16]. We found that the latter interpretation is the most consistent with the observations.

In recent experiments with low-defect small-diameter (~50 nm) Bi nanowires, the result of recent advances in fabrication techniques, we observed that the *h/e* and *h/2e* oscillation becomes very marked and the conductance *G* exhibit steps and plateaus, that we interpret as quantum conductance. We employed the Landauer-Buttiker formalism [17] to find the *mfp* which is found to be very long indicating very high mobility. We found evidence of surface states via transverse *MR* in which we observe the Landau levels of the electron gas at the surface. Since our nanowires are long compared to the *mfp* we can neglect contact resistance and calculate the mobility $\mu$ from *G* and from the surface charge density $\Sigma$. This direct method confirms that mobility is high. This finding is puzzling because the surface states experience surface scattering in detects. Our interpretation is that, because of Lorentz forces, nanowire surface state modes with counterclockwise orbital helicity are spatially displaced away from the surface and therefore decoupled from surface scattering.

The samples used in our experiment have diameters in the range between 45 nm and 55 nm. They were fabricated in several steps, see Supplemental Material [18]. In the first step, the Ulitovsky technique [5] was used to prepare 200-nm wires. This technique involved using a high-frequency induction coil to melt a 99.999% Bi boule within a borosilicate glass capsule while simultaneously softening the glass. Glass fibers containing the 200-nm nanowires were then pulled from the capsule. X-ray and transverse *MR* show that these wires are single crystal. In the second step, fibers containing 200-nm wires were stretched with a capillary puller via the Taylor method. Subsequently the nanowires were annealed at 200 C. Transverse *MR* studies are anisotropic, showing that the wires grown with this method are single crystal or have very few defects [18]. The diameter *d* is estimated from the room temperature resistance and also using scanning electron microscopy (SEM). Electrical contact to the nanowire ends was performed with $In_{0.5}Ga_{0.5}$ solder. The wires are long (fractions of a millimeter) and the contact resistance (of order k$\Omega$) is much less than the zero-field resistance. For 45-55 nm wires, we observed a thermally activated conductivity typical of semiconductors for *T* > 100 K which is different from that of bulk semimetallic Bi. At lower temperatures, the conductivity saturates, signaling that electronic transport is dominated by surface carriers for $T < 10$ K. Some samples exhibit sharp AB oscillation and one sample, with a diameter of 50 nm, was selected for thermopower experiments. In Fig. 1, we present the magnetic field dependence of the derivative of the conductance $G$ with applied field $B$ and thermopower $\alpha$ for a nanowire with diameter $d = 50$ nm and a length $l = 500 \pm 100$ µm at 1.5 K of the sample. $G$ and $\alpha$ measurements are performed together, in the same experimental device, in a 14 T superconducting magnet. $dG/dB$ is calculated from $G$. The low temperature $G$ is $2.8 \times 10^{-6}$ $\Omega^{-1}$. The thermopower is positive indicating that the charge carriers are holes. The inset of Fig. 1 shows a scanning electron microscope (SEM) image of the wire cross-section. Errors in $d$ are 10% because the nanowire is immersed in the glass fiber and the glass charges, thereby interfering with the SEM.

As shown in Fig. 1(a), we observed a deep periodic modulation of the conductance $G$ and thermopower with applied $B$. Two periods with values of $0.98 \pm 0.05$ T and $1.7 \pm 0.1$ T were detected, with the faster oscillation being most visible for $B > \sim 3$ T. A fast Fourier transform (FFT) is shown in Figure 1(b) peaking at 1 and 1.7 T for the slow and fast oscillation respectively. The inset also shows the FFT in the case of a 75 nm nanowire, for comparison purposes. The observed periods can be associated with $h/e$ and $h/2e$ AB effects in 50 nm nanowires considering the experimental errors in the diameter and AB periods. Here the modulation ($\Delta G/G \sim 0.1$) was deep. This is in contrast to previous work, for example $Bi_2Se_3$ nanoribbons [8], in which the modulation was shallow (~0.004). Taking the width of the spectrum at $h/e$ as an estimate for the uncertainty of oscillation frequencies, and assuming the spread in frequencies results from the carriers´ occupation of a region of finite width $w$ near the surface, we obtained an upper bound of ~ 5 nm for $w$. It is estimated that $w$ is less than 5 nm because the spectral width can be augmented by any variation in the wire diameter along its length. The ($B = 0$) low temperature conductance that we observed corresponds to a sheet resistance of 186 $\Omega/\square$. This value is comparable to the room temperature value of 660 $\Omega/\square$ observed in ultrathin (2.5 nm) Bi films that feature high-mobility surface charges [19]. In thin films and small diameter nanowires of bismuth, surface electronic transport dominates the bulk for a number of reasons. Samples of small cross-sections emphasize surface electronic transport. Also, quantum confinement has an effect on bulk carrier densities, depressing the bulk carrier density and further amplifying the importance of the surface bands in the nanowire electronic transport.

We tested the surface via a study of the angular dependence (rotation around the wire axis) and the Shubnikov-de Haas (SdH) oscillations (periodic in $1/B$) of the transverse *MR* (*TMR*), that is, with the field perpendicular to the wire length assigning them to surface states. We considered that the presence of AB oscillations is strong evidence of the two-dimensional character of the surface and therefore we assigned the SdH oscillations to two-dimensional Landau state of the surface carriers. Analysis of the temperature and magnetic field dependence of the SdH oscillations according to Lifshits and Kosevich [20] and Taskin and Ando [21] which is described in detail as Supplemental Material [18] produced $m_\Sigma = 0.25 \pm 0.03$ in units of $m_0$, the electron mass, in good agreement with ARPES [1]. The charge density per unit area $\Sigma$ was estimated from the SdH period $P = 0.06\ T^{-1}$ using $\Sigma = f/(P\phi_0)$, where $f$ is the 2D Landau level degeneracy[4] which is four on account of the two-fold spin degeneracy and two fold helical degeneracy. We find $\Sigma = 1.6 \times 10^{12}/cm^2$ which is less than the ARPES measurement of $8 \times 10^{12}/cm^2$ [22], but not too far off. The 2D Fermi energy $E_F = h^2\Sigma/m_\Sigma$ was found to be 18 meV. Further interpretation of the transverse *MR* is not straightforward. A simple linear fit of the fan diagram gives an intercept of 0.35, an apparently non-trivial value that is unexpected because theoretical analysis indicates that bulk Bi has a trivial TI character [2,3], for which the Berry phase is zero. However, when we modeled the SdH points with a non-linear fit that includes SOC [21], with a $g$-factor of 20, we obtained the expected result; that is, the intercept, and therefore the Berry phase, is zero. However we were not able to independently crosscheck the value of $g$. The analysis in terms of SdH is not straightforward because of the cylindrical geometry of the wires. Treating this as a flat surface perpendicular to the field is not obviously valid since most of the surface is not perpendicular to the magnetic field.

Fig. 2 shows the conductance $G$ of the 50-nm sample at 1.5 K as a function of the magnetic field applied parallel to the wire length. For high fields, $G$ increases by more than one order of magnitude from $B = 0$ to 14 T. The high value of $G$ that is observed at 14 T, $G = 4 \times 10^{-5}\ \Omega^{-1}$, is limited by the contact resistance of 2.5 K$\Omega$. Deep minima, observed for d$G$/d$B$ (Fig. 1), show up as plateaus in $G$. The plateaus in $G$ are shown in the inset for $v = 1$ to 3, where $v$ is the order that can be interpreted as indicating the number of surface conduction channels that are occupied. For large values of $v$, $G$ increases linearly by $G_0 = 3.0 \times 10^{-6}\ \Omega^{-1}$ per unit $v$. Plateaus and the linear dependence of the conductance are fit with a conduction model based on the cyclic opening of one-dimensional channels. According to the Landauer formula [17], $G = G_0\ v$, where $G = (2e^2/h)(mfp/l)$. Our estimate for the *mfp* is 8 μm which is 160 times the wire diameter. $\mu$ can be estimated directly from the nanowire conductance, using $\mu = G/\Sigma$. This estimate depends on $B$ because $G$ increases with magnetic field. For $B = 0$, we find $\mu = 14{,}000\ \text{cm}^2\text{V}^{-1}\text{s}^{-1}$, and for $B \sim 10$ T, we find $\mu = 130{,}000\ \text{cm}^2\text{V}^{-1}\text{s}^{-1}$; this value represents a lower bound, since the measured high field conductance is limited by the contact resistance. These findings confirm the estimate based on Landauer´s expression and indicate that the surface band mobility increases with increasing $B$ for $B < 7$ T, stabilizing at the high value for $B > 7$ T up to the maximum value of $B$ in our measurement. Our mobility estimate is comparable, within order of magnitude, to the $\mu$ reported for suspended graphene [23] (in excess of 600,000 cm$^2$V$^{-1}$s$^{-1}$). In comparison, long graphene nanoribbons also exhibit conductance plateaus in steps of multiples of $e^2/h$ [24,25] associated with quantum confinement subbands.

In previous studies of Bi nanowires and TI nanoribbons, AB effects with both $h/e$ and $h/2e$ flux periods were observed [5-10]. Although the experiments clearly demonstrate the surface origin of the electronic transport, the implication for the mobility and the mechanisms for dissipation were not throughly investigated. The $h/e$ period indicates that the surface charges propagate coherently through the nanowire perimeter. The $h/2e$ period can be associated with the Altshuler-Aronov-Spivak (AAS) oscillations [11,13] that arise in the presence of disorder so that transport in the wire circumference is diffusive. Since our finding of high mobility in Bi nanowires does not support an interpretation of the $h/2e$ period based on disorder, we interpret the observations, including the presence of both $h/e$ and $h/2e$ periods by considering that the dispersion relation of Bi consists of two branches that are spin split and that the transport is quasiballistic ($l > mfp > 2\pi d$). The nanowire 1D subbands' orbital energy $E$ is $E_c (L+He\phi/h)^2 + g\mu_B B m$ where $E_c = h^2/2\pi m_\Sigma d^2$ is the confinement energy. The first and second terms of $E$ are the kinetic and Zeeman energy respectively. $L$ and $m$ are the orbital and spin quantum numbers. The helicity $H$ is $+1$ for clockwise (CW) rotation and $-1$ for counter-CW (CCW) rotation with respect to $B$, respectively. At low temperatures only the subbands with $E < E_F$ are occupied. For $B = 0$, since $E_F = 18$ meV, the subbands with $L$ less than 12 are occupied. With applied field $B$, there are pairs of subband levels crossing the Fermi level ($E = E_F$) per AB cycle, that is, for flux change $\Delta\phi = h/e$. This is illustrated in Fig. 3. Nanowire subbands exhibit a complex behavior that tends to keep the total carrier density constant through the AB cycle . With increasing $B$, $E$ increases for CW subbands and decreases for CCW subbands. With increasing $B$ and constant $E_F$, four Fermi level crossings ($E = E_F$) are found per AB cycle, that is, for $\Delta\phi = h/e$. For each cycle, a pair of CCW states of opposing spin orientations go from occupied to empty, and a pair of CW states with $H = -1$ go from empty to occupied. In nanowires, the opening or closing of a 1D channel causes a peak in $G$. Therefore the presence of pairs of level crossings per cycle explains the presence of both $h/e$ and $h/2e$ periods.

We also observed (see Fig. 2) that $G$ increases with increasing $B$ and $v$. Since the increase of $B$ causes the number of CW modes to decrease and CCW modes to increase, we find that the modes of high mobility are the CCW ones. This effect can be interpreted by considering that the surface charges are holes, the Lorentz force tends to make the orbit of the CW modes closer to the surface of the Bi nanowire while pushing the CCW ones away from the surface, as illustrated in the inset of Fig. 2. CCW modes experience more bulk-like conditions than CW modes. All other factors being equal, closeness to the surface can only decrease mobility via surface scattering. The scattering conditions are reminiscent of the case of high mobility bands in a two-dimensional electron gas field effect transistor [26]. Also, the observed effect is similar to the well-known Chambers´ effect [27]. Its signature is negative $MR$, with $B$ //wirelength, when the cyclotron orbit is smaller than the wire diameter; this has been observed in nanowires with bulk ballistic electronic transport including Na and Bi [28]. We can check if this scenario is reasonable for surface states by roughly estimating the change of orbit diameter $\delta d$ given a Lorentz force $F_L = eV_F BH$. For surface confining potential, we use a harmonic potential $(1/2)k(r-a)^2$ where $r$ is the radius, $a \sim d/2$ where $k$ is estimated considering that $E_F = kw^2$. Then, since $\delta d = F_L/k$, we find that $\delta d = 4$ nm for 10 T. This orbit diameter change is comparable to the width *w* of the interface and therefore this estimate supports our interpretation of the observations where Lorentz forces decouple surface charges from surface scattering.

We have argued that the observed order of magnitude increase of *G* with *B* is a consequence of the increasing number of CCW orbital helical modes of high mobility. Another effect that could potentially cause the increase in *G* is an increase in Fermi energy, which in turn would entail an increase in *Σ* as we increase the field. We ruled this mechanism out by performing the measurements of *α*, which can be paired with the conductance to give $E_F(B)$. Applying the Mott relation [29-31] to the amplitudes, we find that $E_F$ increases by approximately 7 meV for 10 T from the $E_F$ ~ 18 meV without a magnetic field. This result would correspond to a modest 30% increase in *Σ*, which could not explain the observed increase in *G*.

The sensitivity to magnetic fields that we observe suggests that Bi nanowires exhibit strong electric field effects. Rashba bands in BiSe [32] show large sensitivity to electric fields. The magnitude of the effects that we observe in Bi is large as well. Based on our estimate of $E_F(B)$, we deduce that a gate voltage of about 7 meV is equivalent to an applied magnetic field of 9 T, which increases *G* by an order of magnitude. Unlike in the experiments with BiSe where there is a surface-bulk interplay, the response that we observe here is intrinsic to the surface.

In conclusion, we find that 50 nm Bi nanowires exhibit quantized (stepwise) electronic transport under a magnetic field and present unambiguous experimental evidence that Rashba bands exhibit very long mean free paths and high mobilities of around 140,000 $cm^2V^{-1}s^{-1}$ at low temperatures, comparable in order of magnitude to those observed in suspended graphene. Our interpretation is that the high mobility carriers are in select, buried spin-split nanowire orbital helical modes where they avoid surface scattering. Our structured, "thick and buried" surface model is in sharp departure with the flat model previously assumed for surface states.

Our results suggest that spin orbit Bi surface bands can be the basis of quantum materials for future high speed spintronic devices and the basis for further demonstrations of enhanced quantum mechanical electronic transport.

We thank B. Halperin and P. Jarillo-Herrero for their helpful insights. This material is based upon work supported by the National Science Foundation through PREM 1205608 and STC 1231319 . We also acknowledge support by the Boeing Company, the Swiss National Science Foundation, and the Science Technology Center in Ukraine # 5986.

# References

*Corresponding author: T. E. Huber. Address: Howard University, Washington, DC 20059, USA. Tel: 202 806 6768. *E-mail address*: thuber@howard.edu

FIGURE CAPTIONS

FIG. 1. (a) *dG/dB* and thermopower $\alpha$ for a 50-nm wire as a function of *B* along the wirelength at 1.5 K showing minima. The minima order ν is indicated. Inset: SEM cross-sectional image of the (50 ± 5)-nm wire (clear) in its glass envelope (gray background). (b) FFT of the derivative d*G*/d*B* of 50- and 75-nm wires in the entire field range.

FIG. 2. Black solid line. Quantized conductance *G* of the 50-nm Bi nanowire as a function of *B* measured at 1.5 K. The dashed line is the high magnetic fields linear fit $G_0 \nu$ where $G_0 = 3 \times 10^{-6}$ $\Omega^{-1}$. Blue solid line. The scale for *G* has been expanded by 12 so as to make the conductance steps at $\nu = 1, 2$, and 3 more evident. The value of *G* at the steps is indicated. The inset illustrates the nanowire encircled by surface holes in high-mobility CCW (green) and low-mobility CW (red) orbits according to the theoretical model. The confinement potential *V* of surface range *w*, the Lorentz force ($F_{CW}$ and $F_{CCW}$) and orbit radii are also shown.

FIG. 3. Schematic explanation of the data using spin-dependent orbital energies versus magnetic flux, after setting *g* = 2, for 1D subbands calculated using Eq. 1. in the CW (*H* =1) and CCW (*H* = -1) cases. The dashed line represents the Fermi energy. *L* and *L´* are the special orbital quantum numbers that lead to level crossings in the range between a given $\phi$ and $\phi + h/e$. Full and dotted lines indicate occupied and empty levels, respectively.

FIG. 1. Huber et al., (2016).

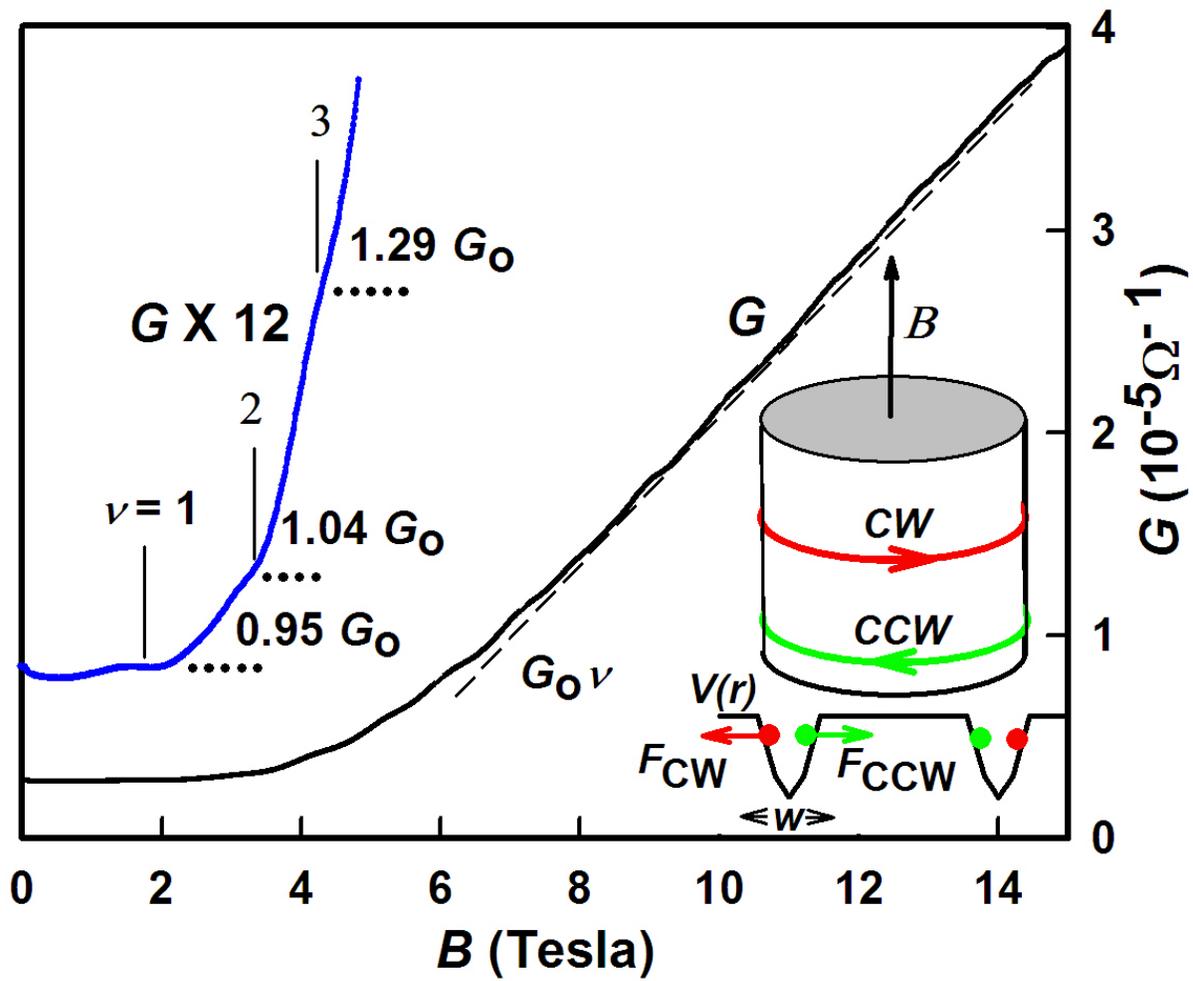

FIG. 2. Huber *et al.*(2016).

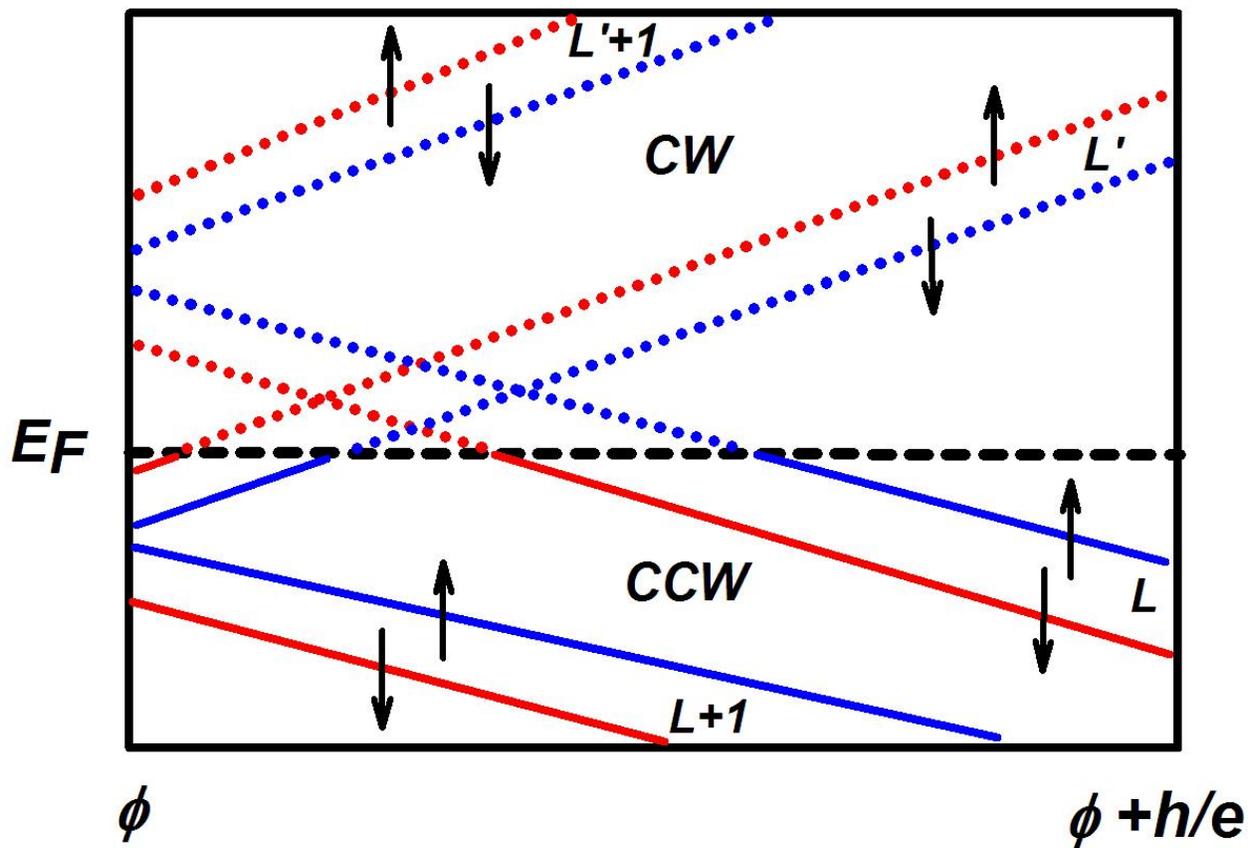

FIG. 3. Huber *et al.* (2016).